\documentclass[twocolumn,showpacs,prc,superscriptaddress]{revtex4}

\usepackage{graphicx}
\usepackage{dcolumn}
\usepackage{bm}
\usepackage{tabularx}

\def\vec#1{\mbox{\boldmath $#1$}}
\newcommand{\refp}[1]{(\ref{#1})}
\newcommand{\Tr}[0]{\mbox{Tr}}
\newcommand{\larw}[1]{\overleftarrow{#1}}
\newcommand{\rarw}[1]{\overrightarrow{#1}}
\newcommand{\del}[2]{\frac{\partial #1}{\partial #2}}
\newcommand{\delr}[1]{\frac{\rarw{\partial}}{\partial #1}}
\newcommand{\dell}[1]{\frac{\larw{\partial}}{\partial #1}}
\newcommand{\ddelr}[1]{\frac{\rarw{\partial^2}}{\partial{#1}^2}}
\newcommand{\ddell}[1]{\frac{\larw{\partial^2}}{\partial{#1}^2}}
\newcommand{\bra}{\langle}
\newcommand{\ket}{\rangle}

\newcommand{\idot}{\!\cdot\!}

\begin{document}


\title{Continuum quasiparticle linear response theory using the Skyrme functional for multipole responses of exotic nuclei}

\author{Kazuhito Mizuyama}
\email{mizu@nt.sc.niigata-u.ac.jp}
\affiliation{Graduate School of Science and Technology, Niigata University, Niigata 950-2181, Japan}

\author{Masayuki Matsuo}%
\affiliation{Department of Physics, Faculty of Science, Niigata University, Niigata 950-2181, Japan}

\author{Yasuyoshi Serizawa}%

\affiliation{Graduate School of Science and Technology, Niigata University, Niigata 950-2181, Japan}

\date{\today}
\begin{abstract} 
 We develop a new formulation of
 the continuum quasiparticle random phase approximation (QRPA)
  in which the velocity dependent terms of the
Skyrme effective interaction are explicitly treated except the
spin dependent and the Coulomb terms.  
Numerical analysis using the
SkM$^*$ parameter set is performed
for the isovector dipole and the isovector/isoscalar
 quadrupole responses in $^{20}$O and $^{54}$Ca. 
It is shown that the energy-weighted sum rule including the
enhancement factors for the isovector responses is satisfied with
good accuracy. We investigate also how the velocity dependent 
terms influence the strength distribution and the transition
densities of the low-lying surface modes and the giant 
resonances.
\end{abstract}

\maketitle

\section{\label{sec:intro}Introduction}

Nuclei near the neutron drip-line provide us with many
new physics issues which arise from the
presence of weakly bound neutrons and the coupling to
unbound neutron states. The ground
state and the excitation modes of a near-drip-line nucleus
are indeed very different from those of stable nuclei
as is testified by the observations of
the neutron halo\cite{tanihata}, the neutron skin\cite{ozawa} 
and the soft dipole excitation\cite{aumann-review}.
In addition the nucleon correlations such as the pairing 
may also be influenced in the new 
circumstances\cite{dobaczewski,dobaczewski96}.
Consequently there has been
considerable efforts in the last 
two decades to develop nuclear many-body theories
toward this direction.

Focusing on near-drip-line nuclei in the medium mass region,
theoretical approaches
based on the self-consistent mean-field methods or
the density functional theories are of great promise.
The Hartree-Fock-Bogoliubov (HFB) theory \cite{ringschuck},
especially those employing the coordinate-space representation 
\cite{dobaczewski,dobaczewski96,bulgac-hfb},
has been playing a central role to describe the ground state and the pair
correlation.
The HFB theory provides us also with the basis for
further theoretical developments to describe the dynamics, e.g.
the excitation modes built on the ground state. Indeed new schemes of the
quasiparticle random phase approximations (QRPA) 
formulated on the basis of the
coordinate-space HFB have been recently proposed and applied extensively
to studies of multipole responses of unstable nuclei
\cite{engel,bender,matsuo01,matsuo02,matsuo-mizu-seri,matsuo06,
khan,khan2,khan3,khan-goriely,paar,paar2,yamagami2,yamagami-aizu,yamagami3,
terasaki,terasaki2,yoshida} (see also references in Ref.\cite{paaretal}).

There are two important requirements to be considered when the
HFB+QRPA theories are applied to near-drip-line nuclei. 
First of all, the coupling of excitation modes 
to the continuum states have to be taken into account since
most of excitation modes including even the low-lying excitations
are located near or above the nucleon separation energy. This can
be achieved by means of the continuum QRPA methods 
\cite{matsuo01,matsuo02,matsuo-mizu-seri,matsuo06,khan,khan2}.
Secondly, the QRPA description should be consistent with the
HFB description of the ground state in the sense that the
same effective interaction or the same density functional
should be used for both descriptions. 
If this is achieved, one can calculate the ground and excited states 
solely from the effective interaction (or
the energy density functional) without relying on
phenomenological parameterization of the mean-fields.
This is often called the requirement of the self-consistency.
The two requirements, however, have been 
in a trade-off relation in the actual implementations.
Namely in the continuum 
QRPA methods which fulfill the first requirement the self-consistency has been
left behind since the residual
interaction for the QRPA description is often approximated to a tractable
simple contact force\cite{matsuo01,matsuo02,matsuo-mizu-seri} 
or the Landau-Migdal forces \cite{khan,khan2,matsuo06}. 
On the other hand,
recently developed fully self-consistent QRPA's using 
the Skyrme functional\cite{terasaki,terasaki2} and
the relativistic mean-field functional\cite{paar,paar2} treat
approximately the continuum states by employing
the finite-box discretization or the discrete oscillator
basis.

It is therefore important to develop a new formulation of the
continuum QRPA which is based on the nuclear density functional
and thus satisfies the self-consistency as precisely as possible.
In the present paper we try to make a one step progress in this direction.

To this end we shall proceed in the following way.
We start with the Skyrme's Hartree-Fock energy functional 
combined with the pair correlation energy. 
We then use this functional not only for the
static Hartree-Fock-Bogoliubov mean-fields but also 
to derive the residual interaction to be used
in the continuum QRPA. In formulating the new continuum QRPA,
we pay special attention to the energy weighted sum rule,
which is not satisfied in the previous continuum QRPA's
\cite{matsuo01,matsuo02,matsuo-mizu-seri,matsuo06,khan,khan2}. 
To satisfy this we take into account explicitly
the velocity dependent central terms of the Skyrme effective
interaction to derive the residual interaction, and then 
implement the residual interaction 
into the Green's function formulation of the continuum QRPA
proposed in Ref.\cite{matsuo01}. 
In the present paper, however, we do not include the spin dependent
densities and the Coulomb interaction in deriving the residual interaction,
and hence the goal of the full self-consistency is not achieved yet.
Our formulation of the Skyrme QRPA is similar to that of 
Ref.\cite{yamagami2,yamagami-aizu}
apart from the treatments of the continuum quasiparticle states,
on which we impose the out-going wave boundary condition instead
of the finite-box discretization.

By performing numerical calculations, we shall demonstrate 
that the Skyrme continuum QRPA  in the present formulation  indeed satisfies the sum rule
  as far as the dipole and quadrupole responses with natural parities
are concerned.  
We shall also show  that 
the inclusion of the velocity dependent terms 
gives better description of the strength function and
the transition densities of the multipole responses
in comparison with the previous continuum QRPA that utilizes 
residual interactions of the simple contact forces.

The paper is organized as follows. 
The formulation is given in the next section. In Section III
we present results of numerical calculation performed
 for
 the isovector dipole 
and the isoscalar/isovector quadrupole responses in neutron-rich 
O and Ca isotopes. We discuss both the low-lying excitations
and the giant resonances.
We shall illustrate in 
detail the importance of taking account of the velocity dependent
terms by comparing with the Landau-Migdal approximation of the
Skyrme effective   interaction\cite{LM-Skyrme,GiaiSagawa}.  
The conclusions are drawn in Section IV.

\section{\label{sec:cqrpa} Continuum QRPA using the Skyrme functional}

In this section, we give a formulation of the continuum QRPA which is
based on the Skyrme functional.

We start with the energy functional of the system
defined for a determinantal many-body state vector $\left|\Phi(t)\right\ket$
of the generalized form in which the pair correlation is taken into
account by means of the Bogoliubov's quasiparticle
method\cite{ringschuck}. 
The time-dependence is explicitly written here since we consider 
dynamical multipole responses of the system under a time-dependent
perturbation. 
The energy functional 
$E=E_{Skyrme} + E_{pair}$ consists
of the Skyrme Hartree-Fock energy $E_{Skyrme}$ and the pair correlation energy
$E_{pair}$. 
The Skyrme Hartree-Fock energy $E_{Skyrme}$ is expressed in terms
of the local density $\rho_q(\vec{r},t)$, its spatial derivatives
$\vec{\nabla}\rho_q(\vec{r},t)$ and $\Delta\rho_q(\vec{r},t)$, the
current density $\vec{j}_q(\vec{r},t)$, the kinetic energy density
$\tau_q(\vec{r},t)$, the spin density $\vec{s}_q(\vec{r},t)$
and the spin-orbit tensor 
$\vec{J}_{q}(\vec{r},t)$ where $q=n,p$ stands for the neutron or proton
components\cite{bonche,bender-rev}. 
Given a parameter set such as SIII\cite{SIII}, SkM*\cite{SkMs} and
SLy4\cite{SLy4}, the Skyrme Hartree-Fock energy functional 
$E_{Skyrme}[\rho,\vec{\nabla}\rho,\Delta\rho,\tau,\vec{j},\vec{s},\vec{J}]$ 
is completely specified.
Concerning the pair correlation energy $E_{pair}$, 
we use the one evaluated for the 
the density-dependent delta 
interaction (DDDI) \cite{bertsch,dobaczewski-pair}
\begin{eqnarray}
V_{pair}(1,2)
=
\frac{1}{2}V_0(1-P_\sigma)
\left[1-\eta\left(\frac{\rho(\vec{r})}{\rho_c}\right)^\gamma \right]
\delta(\vec{r}-\vec{r}').
\end{eqnarray}
$E_{pair}$ is a functional 
of the local density $\rho_q(\vec{r},t)$  and the local pair densities
\begin{equation}
\tilde{\rho}_{\pm q}(\vec{r},t)
=\left\langle \Phi (t)\right| \psi_q^\dagger(\vec{r}\downarrow)\psi_q^\dagger(\vec{r}\uparrow)
\pm \psi_q(\vec{r}\uparrow)\psi_q(\vec{r}\downarrow)\left| \Phi (t)\right\rangle.
\end{equation}

Application of the static variational principle to the total energy functional 
$E_{Skyrme} + E_{pair} $ leads to the Hartree-Fock-Bogoliubov equation
\begin{equation}
\mathcal{H}_{0q}\phi_q(\vec{r}\sigma) = E_q\phi_q(\vec{r}\sigma)
\end{equation}
for the quasiparticle wave function
\begin{equation}
\phi_q(\vec{r}\sigma)
=
\left(
\begin{array}{c}
\varphi_{q,1}(\vec{r}\sigma) \\
\varphi_{q,2}(\vec{r}\sigma)
\end{array} 
\right).
\end{equation}
Here 
\begin{eqnarray}
\mathcal{H}_{0q}=
\left(
\begin{array}{cc}
h_q-\lambda_q & \tilde{h}_q \\
\tilde{h}_q^* & -h_q^*+\lambda_q
\end{array}
\right)
\end{eqnarray}
is the 2 $\times$ 2 matrix representation of the HFB mean-field Hamiltonian
\begin{eqnarray}
\hat{h}
&=&\sum_q\int d\vec{r} d\vec{r}' \sum_{\sigma,\sigma'} h_q(\vec{r}\sigma,\vec{r}'\sigma')
\psi^\dagger_q(\vec{r}\sigma) \psi_q(\vec{r}'\sigma') 
\nonumber\\
&&+ 
{1 \over 2}\int d\vec{r} d\vec{r}' \sum_{\sigma,\sigma'}\tilde{h}_q(\vec{r}\sigma,\vec{r}'\tilde\sigma')
\psi^\dagger_q(\vec{r}\sigma) \psi^\dagger_q(\vec{r}'\tilde{\sigma}')  
\nonumber\\
&&
+ h.c.
\end{eqnarray}
The Hartree-Fock Hamiltonian $h_q$ and the pair potential $\tilde{h}_q$ 
are defined through
the functional derivative of 
the energy functionals $E_{Skyrme}$ and $E_{pair}$, respectively. 

We consider multipole response of the nucleus under a small
time-dependent external perturbation 
\begin{eqnarray}
\hat{V}_{ext}(t) 
&=& 
e^{-i\omega t}\sum_q\int d\vec{r}f_q(\vec{r})
\sum_{\sigma}\psi_q^\dagger(\vec{r}\sigma)\psi_q(\vec{r}\sigma)
\nonumber\\
&&
+ h.c.
\end{eqnarray}
expressed in terms of a one-body spin-independent local field 
$f_q(\vec{r})$, for which we take a multipole field $\propto r^{L}Y_{LM}$ such as the
electric dipole and the isoscalar/isovector quadrupole fields.

The external perturbation causes the induced fields in the Hartree-Fock
mean-field and the pair potential, which we denote $\delta h_q$ and
$\delta \tilde{h}_q$, respectively. $\delta h_q$ and $\delta \tilde{h}_q$
are expressed in terms of fluctuations in
the various one-body densities 
\begin{eqnarray}
&&\delta\rho_q(\vec{r},t),   \delta\vec{\nabla}\rho_q(\vec{r},t), \delta\Delta\rho_q(\vec{r},t),
\delta\tau_q(\vec{r},t), \delta\vec{j}_q(\vec{r},t), \nonumber\\
&&\delta \vec{J}_q(\vec{r},t), \delta \vec{s}_q(\vec{r},t), \nonumber\\
&&\delta\tilde{\rho}_{\pm q}(\vec{r},t)
\label{eq:denstfluc}
\end{eqnarray}
and the second derivatives of the energy functional.
A fully self-consistent QRPA based on the Skyrme HFB functional can be 
constructed if one considers all the kinds of density fluctuations in Eq.(\ref{eq:denstfluc}).
In the previous continuum QRPA approaches, however, only the fluctuations in the 
local densities $\delta\rho_q(\vec{r},t)$ and $\delta\tilde{\rho}_{\pm q}(\vec{r},t)$,
and the induced fields associated with these
density fluctuations have been taken into account
\cite{matsuo01,matsuo02,matsuo-mizu-seri,matsuo06,khan,khan2}. 
Although this approximation has a large number of
practical usefulness, it is not sufficient in some respects: it violates
the energy weighted sum rule when the Skyrme HF mean-field 
with the effective mass is adopted. This is because 
the current conservation law is not satisfied when 
the velocity dependent parts (the terms proportional to the $t_1$ and $t_2$ 
terms) of the Skyrme interaction and the current fluctuations $\delta\vec{j}_q$
are neglected in the RPA\cite{tsai,suzuki,BM2,stringari}. We aim at 
improving this point. 
For this purpose we shall include
all the 
density fluctuations that are responsible for the energy weighted
sum rule for  
the responses caused by the spin independent local multipole fields.
They are 
the fluctuations $\delta\vec{j}_q$, $\delta\vec{\nabla}\rho_q$, 
 $\delta\Delta\rho_q$, and $\delta\tau_q$ in the current,
the  spatial derivatives of the density and the kinetic energy density. 
It is more preferable to take into account also the fluctuations in the spin-dependent densities
$\vec{s}_q$ and $\vec{J}_q$, but we neglect them in the present work.
This is one approximation which remains 
in the present approach. 
We neglect the residual Coulomb interaction. 
The cross derivatives 
among $\rho_q$ and $\tilde{\rho}_{\pm q}$ are also neglected.
These are the second approximation 
we introduce. 
Consequently the full self-consistency is not fulfilled,
but the  treatment of the residual interaction   is significantly improved compared to the 
previous continuum QRPA approaches
\cite{matsuo01,matsuo02,matsuo-mizu-seri,matsuo06,khan,khan2}
in the sense that 
the present formalism allows us to describe 
the correct energy weighted sum rule for the multipole responses.
Note also that the approximate treatment of the particle-hole 
residual interaction is comparable to that adopted in 
the currently available continuum RPA approaches which 
utilize the Skyrme-Hartree-Fock functional without taking into the pairing
\cite{LIU,hamamoto1,hamamoto2,sagawa}.   On the other hand,
we should keep in mind that the approximation neglecting
the spin dependent densities may not be justified for multipole
responses with unnatural parities involving spin excitations.   

The induced fields under the above approximations are expressed as 
\begin{eqnarray}
&&
\delta h_q 
=
\sum_{q'}a_{qq'}\delta\rho_{q'}+b_{qq'}\delta\Delta_+\rho_{q'}
+
\left[\larw{\Delta}+\rarw{\Delta}\right]b_{qq'} \delta\rho_{q'}
\nonumber\\
&&\hspace{0.5cm}
+
\left[\larw{\vec{\nabla}}+\rarw{\vec{\nabla}}\right]\idot c_{qq'} \delta\vec{\nabla}\rho_{q'}
+
\left[\larw{\vec{\nabla}}-\rarw{\vec{\nabla}}\right]\idot b_{qq'} 
\delta 2i\vec{j}_{q'}
\label{eq:resinthf}
\end{eqnarray}
and
\begin{equation}
\delta \tilde h_q =\tilde{a}_q\delta\tilde{\rho}_{+q} - \tilde{a}_q\delta\tilde{\rho}_{-q}.
\label{eq:resintpair}
\end{equation}
The functions 
$a_{qq'}, b_{qq'}, c_{qq'}$ and $\tilde{a}_q$ are 
expressed
in terms of the local densities and the effective interaction 
parameters. The definition
of $a_{qq'},b_{qq'}$ and $c_{qq'}$ follows Ref.~\cite{sagawa}, and their detailed
expressions are given in Appendix.
Note here that  we  use
$\delta\Delta_+\rho_q(\vec{r},t)\equiv 
(\Delta+\Delta')\delta\rho_q(\vec{r}\vec{r'},t)|_{\vec{r'}=\vec{r}}
= \delta\Delta\rho_q(\vec{r},t)-2\delta\tau_q(\vec{r},t)$ 
in place of $\delta\Delta\rho_q(\vec{r},t)$, where $\rho_q(\vec{r}\vec{r'},t)$
is the density matrix. We also attached a factor $2i$ to the current fluctuation
$\delta\vec{j}_q$ so that
$\delta 2i\vec{j}_q=(\vec{\nabla}-\vec{\nabla}')
\delta\rho_q(\vec{r}\vec{r'},t)|_{\vec{r'}=\vec{r}}$
becomes in parallel with $\delta\vec{\nabla} \rho_q=(\vec{\nabla}+\vec{\nabla}')
\delta\rho_q(\vec{r}\vec{r'},t)|_{\vec{r'}=\vec{r}}$.
 The fluctuation in the kinetic energy density 
$\delta\tau_q(\vec{r},t)$ does not appear here since it
 can be eliminated from the induced field by a partial integration. Nevertheless
we eventually include $\delta\tau_q$ as explained later.

The induced fields are also represented in the 2 $\times$ 2 matrix
form as 
\begin{equation}  
\left(
\begin{array}{cc}
\delta h_q & \delta\tilde{h}_q \\
\delta\tilde{h}_q^* & -\delta h_q^*
\end{array}
\right)
=
 \sum_{\beta}
  \mathcal{B}^\beta \hat{O}^{\beta}
  \sum_\gamma \kappa_{\beta\gamma}\delta\rho_{\gamma}
\label{eq:resintmat}
\end{equation}
where $\delta\rho_\gamma$ is a collective
notation for 
\begin{equation}
\delta\rho_\gamma \in 
\delta\rho_q,
\delta\Delta_+\rho_q,\delta\vec{\nabla}\rho_q,\delta 2i\vec{j}_q,\delta\tilde{\rho}_{\pm q}.
\end{equation}
and $\hat{O}^{\beta}$ denotes the derivative operators 
$1, \larw{\Delta}+\rarw{\Delta},
\larw{\vec{\nabla}}-\rarw{\vec{\nabla}}$
and $\larw{\vec{\nabla}}+\rarw{\vec{\nabla}}$ while 
$\mathcal{B}^\beta$ stands for one of the 2$\times$2 matrices
$
\left(
\begin{array}{cc}
1 & 0 \\
0 & -1
\end{array}
\right), 
\left(
\begin{array}{cc}
1 & 0 \\
0 & 1
\end{array}
\right), 
\left(
\begin{array}{cc}
0 & 1 \\
1 & 0
\end{array}
\right) \mbox{and }
\left(
\begin{array}{cc}
0 & 1 \\
-1 & 0
\end{array}
\right).
$
$\kappa_{\beta\alpha}$ represents the functions
$a_{qq'}, b_{qq'}, c_{qq'}$ and $\tilde{a}_q$
in  Eqs.(\ref{eq:resinthf}) and (\ref{eq:resintpair}). 
The correspondence among $\hat{O}^\alpha$, $\cal{B}^\alpha$ $\kappa_{\beta\gamma}$ 
and $\delta\rho_\gamma$ 
is shown in Table \ref{tb:optable}.

The external perturbation and the induced fields 
cause quasi-particle excitations, which in turn bring about 
fluctuations in the densities 
$\delta\rho_q,\delta\tilde{\rho}_{\pm q},
\delta\Delta_+\rho_q,\vec{\nabla}\delta\rho_q$
and $\delta 2i\vec{j}_q$. This relation
is given by the linear response equation, which is written as
\begin{widetext}
\begin{equation}
\delta\rho_\alpha(\vec{r},\omega) = \sum_\beta \int d\vec{r}' 
R_{0q}^{\alpha\beta}(\vec{r}\vec{r}',\omega)\left[\sum_\gamma\kappa_{\beta\gamma}(\vec{r}')\delta\rho_\gamma(\vec{r}',\omega)
+ \delta_{\beta, 0q}f_q(\vec{r})\right]
\label{eq:lineq-cqrpa}
\end{equation}
in the frequency domain.
\end{widetext}
Here $R_{0q}^{\alpha\beta}(\vec{r}\vec{r}',\omega)$ is the unperturbed response function for the density
$\delta\rho_\alpha$ and the field $\mathcal{B}^\beta\hat{O}^\beta$. 
Using the Green's function formalism of the continuum QRPA\cite{matsuo01}, 
the unperturbed
response function is expressed as 
\begin{widetext}
\begin{eqnarray}
R_{0q}^{\alpha\beta}(\vec{r}\vec{r}',\omega)
&=&
{1 \over 4\pi i}
\int_C dE
\Tr
\left[
\mathcal{A}^\alpha
\hat{O}^{\alpha}(\vec{r})
\mathcal{G}_{0q}(\vec{r}\vec{r}',E+\hbar\omega+i\epsilon)
\mathcal{B}^\beta
\hat{O}^{\beta}(\vec{r}')
\mathcal{G}_{0q}(\vec{r}'\vec{r},E)
\right]\nonumber\\
&&
+
{1 \over 4\pi i}
\int_C dE
\Tr
\left[
\mathcal{A}^\alpha
\hat{O}^{\alpha}(\vec{r})
\mathcal{G}_{0q}(\vec{r}\vec{r}',E)
\mathcal{B}^\beta
\hat{O}^{\beta}(\vec{r}')
\mathcal{G}_{0q}(\vec{r}'\vec{r},E-\hbar\omega-i\epsilon)
\right]
\label{eq:resfunc}
\end{eqnarray}
\end{widetext}
in terms of the quasi-particle Green's function
$\mathcal{G}_{0q}(E)=(E-\mathcal{H}_{0q})^{-1}$ and a contour integral in the 
complex energy plane. 
The complex energy integral is performed on a rectangular contour $C$ enclosing
the negative energy part of the real $E$ axis with the two sides located at $\pm {i \over 2}\epsilon$
\cite{matsuo01}.
Here $\epsilon$ is a small parameter which plays a role of the smoothing
energy width.
The matrices $\mathcal{A}^\alpha$ 
and $\mathcal{B}^\beta$ and
the operators $\hat{O}^\alpha$  and $\hat{O}^\beta$ 
follow Table \ref{tb:optable}, but we remark that
the matrix $\mathcal{A}^\alpha$ takes a form 
$
\mathcal{A}^\alpha= 
\left(
\begin{array}{cc}
2 & 0 \\
0 & 0
\end{array}
\right)
$
for the particle-hole densities 
$\delta\rho_q, \delta\Delta_+\rho_q, \delta\vec{\nabla}\rho_q,\delta 2i \vec{j}_q$ and $\delta\tau_q$
 while the matrix $\mathcal{B}^\beta$  has the following
definitions:  
$
\mathcal{B}^\beta= 
\left(
\begin{array}{cc}
1 & 0 \\
0 & -1
\end{array}
\right)
$
for the   'time-even' quantities $\delta\rho_q, \delta\Delta_+\rho_q$ and 
$\delta\vec{\nabla}\rho_q$,  and 
$
\mathcal{B}^\beta=
\left(
\begin{array}{cc}
1 & 0 \\
0 & 1
\end{array}
\right)
$ for the 'time-odd'  $\delta 2i \vec{j}_q$ (See Table \ref{tb:optable}).

Let us assume 
the spherical symmetry of the ground state, and we
apply the multipole decompositions.
Let $L$ be the multipolarity of the excitation modes under 
consideration. The fluctuation in the scalar quantities
$\delta\rho_\alpha=\delta\rho_q, \delta\tilde{\rho}_{\pm q}$ and $\delta\Delta_+\rho_q$
are expanded as 
\begin{equation}
\delta\rho_\alpha(\vec{r},\omega) = Y_{LM}(\hat{\vec{r}})[\delta\rho_{\alpha}]_L/r^2,
\end{equation}
and we now consider only the radial functions
$[\delta\rho_{\alpha}]_L = 
[\delta\rho_{q}]_L, [\delta\tilde{\rho}_{\pm q}]_L$ and 
$[\delta\Delta_+\rho_q]_L$. 
Concerning the vector quantities $\delta\vec{\rho}_\alpha=\delta\vec{\nabla}\rho_q$ and $\delta 2i\vec{j}_q$, they
are  expanded as
\begin{equation}
\delta\vec{\rho}_\alpha(\vec{r},\omega) = 
\sum_{\lambda=L\pm1}\vec{Y}_{L\lambda M}(\hat{\vec{r}})[\delta\vec{\rho}_{\alpha}]^\lambda_L/r^2,
\end{equation}
in terms of the vector spherical harmonics $\vec{Y}_{L\lambda M}$
and the radial functions 
$[\delta\vec{\rho}_{\alpha}]^\lambda_L
= [\delta\vec{\nabla}\rho_q]_{L}^{\lambda=L\pm 1}$ and
$[\delta 2i \vec{j}_q]_{L}^{\lambda=L\pm 1}$. 
Note that only the terms 
$\lambda=L\pm 1$ remain here 
since we consider the multipole excitations with the natural parity.
Then the
linear response equation 
(\ref{eq:lineq-cqrpa}) is rewritten as an equation for 
the relevant radial functions
$[\delta\rho_{q}]_L, [\delta\tilde{\rho}_{\pm q}]_L,
[\delta\Delta_+\rho_q]_L, [\delta\vec{\nabla}\rho_q]_{L}^{\lambda=L\pm 1}$ and
$[\delta 2i \vec{j}_q]_{L}^{\lambda=L\pm 1}$. Denoting collectively 
these density fluctuations $\delta\rho_{\alpha L}$, 
the linear response equation for these variables is given by
\begin{eqnarray}
&&\delta\rho_{\alpha L}(r,\omega)
\nonumber\\
&&
=\sum_\beta\int dr' R_{0,qL}^{\alpha\beta}(rr'\omega)
\nonumber\\
&&\hspace{1cm}\times
\left[\sum_\gamma \kappa_{\beta\gamma}\delta\rho_{\gamma L}(r',\omega)/r'^2+\delta_{\beta,0q}f_{qL}(r')\right]
\nonumber\\
\label{eq:lineq}
\end{eqnarray}
using the  unperturbed response function for the fixed multipolarity $L$
\begin{widetext}
\begin{eqnarray}
&&
R^{\alpha\beta}_{0,qL}(rr',\omega)\nonumber
\\
&=&
\frac{1}{4\pi i}
\int_C dE
\sum_{ljl'j'}
\frac{|\bra l'j'||Y_L||lj\ket|^2}{2L+1}
\Tr
\left[
\mathcal{A}^\alpha
\hat{O}^{\alpha}_{ljl'j'}(r)
\mathcal{G}_{0,ql'j'}(rr',E+\hbar\omega+i\epsilon)
\mathcal{B}^\beta
\hat{O}^{\beta}_{l'j'lj}(r')
\mathcal{G}_{0,qlj}(r'r,E)
\right]\nonumber\\
&&+
\Tr
\left[
\mathcal{A}^\alpha
\hat{O}^{\alpha}_{l'j'lj}(r)
\mathcal{G}_{0,qlj}(rr',E)
\mathcal{B}^\beta
\hat{O}^{\beta}_{ljl'j'}(r')
\mathcal{G}_{0,ql'j'}(r'r,E-\hbar\omega-i\epsilon)
\right].
\nonumber\\
\label{eq:resfunc2}
\end{eqnarray}
\end{widetext}
Here $\mathcal{G}_{0,qlj}(r'r,E)$ is the $2\times 2$ radial 
HFB Green's function for 
specified orbital and total angular momenta $l$ and $j$, and
$\hat{O}^{\beta}_{l'j'lj}$ is the radial derivative operator corresponding
to the previously defined $\hat{O}^{\beta}$. 
Their explicit forms are given 
in Table \ref{tb:optable}.
We adopt the exact form for the radial HFB Green's function\cite{Belyaev} 
constructed as
\begin{widetext}
\begin{equation}
\label{eq:greenex}
\mathcal{G}_{0,qlj}(rr',E)=\sum_{s,s'=1,2}
c_{qlj}^{ss'}(E)\left(
\theta(r-r')\phi_{qlj}^{(+s)}(r,E){\phi_{qlj}^{({\rm r}s') T}}(r',E)
+\theta(r'-r)\phi_{qlj}^{({\rm r}s')}(r,E){\phi_{qlj}^{(+s) T}}(r',E)
\right)
\end{equation}
\end{widetext}
in terms of two independent solutions
$\phi_{qlj}^{({\rm r}s)}(r,E) (s=1,2)$ regular at the origin $r=0$ of
the radial HFB equation
and two independent 
solutions $\phi_{qlj}^{(+s)}(r,E) (s=1,2)$ satisfying
the out-going boundary condition. 
(The construction (\ref{eq:greenex}) is the same as that used 
in Refs.\cite{matsuo01,Belyaev}
except that the effective mass should be taken into account in the definitions
of the Wronskian and the coefficients $c_{qlj}^{ss'}$
while in Refs.\cite{matsuo01,Belyaev} the bare mass is assumed.).
In this way the exact treatment of the continuum single-particle states satisfying
the proper boundary condition is
implemented in the QRPA formalism.

Note that in Table \ref{tb:optable} we use the following convention for the derivative operators 
marked with the right/left-sided  arrows
such as  $\dell{r} \pm \delr{r}$.
When this derivative  is inserted in
$\hat{O}^{\beta}_{l'j'lj}(r')$ in the first term of r.h.s. of Eq.(\ref{eq:resfunc2}), the 
derivative symbol $\delr{r'}$ with the right-sided arrow indicates that
it acts on the coordinate $r'$ in the Green's function
$\mathcal{G}_{0,qlj}(r'r,E)$ while the other one 
$\dell{r'}$ with the left-sided arrow acts
on the Green's function 
$\mathcal{G}_{0,ql'j'}(rr',E+\hbar\omega+i\epsilon)$.
The same rule
is applied also to the operator $\hat{O}^{\alpha}_{ljl'j'}(r)$, i.e., 
$\delr{r}$ acts on $r$ in $\mathcal{G}_{0,ql'j}(rr',E+\hbar\omega+i\epsilon)$
while $\dell{r}$ on $r$ in $\mathcal{G}_{0,qlj}(r'r,E)$. 
The ordering of the operators and the Green's functions makes sense in Eq.(\ref{eq:resfunc2}).
    
To obtain a numerical solution of the linear response
equation, we need to rewrite further Eq.(\ref{eq:lineq}).
When the radial derivative operators $\del{}{r}$ and $\del{}{r'}$ 
act on the radial HFB Green's function  like 
$\delr{r}\mathcal{G}_{0,qlj}(rr',E)\dell{r'}$,  a singular term proportional to
$\frac{2m^*_q(r)}{\hbar^2}\delta(r-r')$ emerges. We need to treat these
singular terms separately in the numerical calculation. For this purpose
we rewrite the derivative of the Green's function into 
 singular and regular parts
\begin{eqnarray}
\delr{r}\mathcal{G}_{0,qlj}(rr',E)\dell{r'}
&=&
-\frac{2m^*_q(r)}{\hbar^2}\delta(r-r')\nonumber\\
&&+
\widetilde{\delr{r}}\mathcal{G}_{0,qlj}(rr',E)\widetilde{\dell{r'}}
\label{eq:divex1}
\end{eqnarray}
where the regular part (the second term in r.h.s denoted with the
tildered derivatives) is defined as 
a part of $\delr{r}\mathcal{G}_{0,qlj}(rr',E)\dell{r'}$ that arises from
the action of the derivatives on the wave functions 
$\phi_{lj}^{({\rm r}s)}$ and $\phi_{lj}^{(+s)}$ in Eq.(\ref{eq:greenex}), but not on the 
Heaviside theta function
$\theta(r-r')$. 
Inserting this decomposition into the response
function(Eq.(\ref{eq:resfunc2})), the r.h.s. of the linear response equation
(\ref{eq:lineq})
is decomposed into two parts:
\begin{widetext}
\begin{eqnarray}
\delta\rho_{\alpha L}(r\omega)
&&
=
\sum_\beta\int \!\!dr' \tilde{R}_{0,qL}^{\alpha\beta}(rr'\omega)
\!\!\left[\sum_\gamma \kappa_{\beta\gamma}(r')\delta\rho_{\gamma L}(r'\omega)/r'^2+\delta_{\beta,0q}f_{qL}(r')\right]
\nonumber\\
&&
+
2
\sum_{\beta}
S^{\alpha\beta}_q(r)
\left[\sum_\gamma
\tilde{\kappa}_{\beta\gamma}(r)\delta\rho_{\gamma L}(r\omega)/r^2
+\delta_{\beta,0q}f_{qL}(r)
\right]
\label{eq:lineq2}
\end{eqnarray}
where
\begin{equation}
\delta\rho_{\alpha L}\in 
[\delta\rho_q]_{L},
[\delta\Delta_+\rho_q]_L,[\delta\vec{\nabla}\rho_q]_L^{\lambda=L\pm1},
[\delta\tau_q]_{L},[\delta 2i\vec{j}_q]_L^{\lambda=L\pm1},
[\delta\tilde{\rho}_{\pm q}]_L.
\end{equation}
\end{widetext}

Here $\tilde{R}_{0,qL}^{\alpha\beta}$
denotes a part of the response function which contains only the
regular parts of the derivatives of $\mathcal{G}_{0,qlj}$.  
Its expression is the same as that of $R_{0,qL}^{\alpha\beta}$(Eq.\refp{eq:resfunc2})
except 
that the derivatives of the Green's function, e.g. 
$\delr{r}\mathcal{G}_{0,qlj}(rr',E)\dell{r'}$ in Eq.(\ref{eq:divex1}) is replaced by the corresponding regular part 
$\widetilde{\delr{r}}\mathcal{G}_{0,qlj}(rr',E)\widetilde{\dell{r'}}$.
On the other hand, the second term
of r.h.s. of Eq.(\ref{eq:lineq2}) represents
 contribution from the singular terms such as $\frac{2m^*_q(r)}{\hbar^2}\delta(r-r')$. 
The integral $\int dr'$ disappears in this term because of the delta
function. 
The expressions of $S_q^{\alpha\beta}$ are given 
in Appendix.
Note that 
$S_q^{\alpha\beta}$ is a one-point function 
independent of the frequency $\omega$,
expressed in terms of local quantities such as
$\rho_q(r)$, $\tau_q(r)$, $m^*_q(r)$ and their derivatives. 

It is noted that the linear response equation (\ref{eq:lineq2}) includes the fluctuation $[\delta \tau_q]_L$ 
in the kinetic energy density
$\tau_q$ as a dynamical variable to be considered. 
This is because $[\delta \tau_q]_L$ emerges from
the singular terms associated with 
the linear response equation for $[\delta \Delta_+\rho_q]_L$.
Finally we make a little remark on the structure of the singular terms. 
The presence of the singular terms has been notified in the formulation
of the Skyrme-HF plus continuum
RPA\cite{sagawa,LIU}  where the pairing is neglected. In the present Skyrme-HFB plus 
continuum QRPA approach, the structure of the singular terms is more
involved since the response function contains two single-particle 
HFB Green's functions
(instead of one Green's function
 in the case of the continuum RPA\cite{sagawa,LIU}).
Looking at the expression of Eq.(\ref{eq:resfunc2}), it may appear that
 products of two delta functions $\frac{2m^*_q(r)}{\hbar^2}\delta(r-r')$ emerge from the singular terms of
two HFB Green's functions in Eq.(\ref{eq:resfunc2}). 
Such a term however does not contribute to the response
function since it has no energy dependence and hence it vanishes
when the contour integral in the complex energy plane is performed. 

\begin{widetext}
\begin{center}
\begin{table}
\begin{tabular}{c c c c c}
\vspace{0.25cm}
\\
\hline
$\delta\rho_\alpha$ & 
$\cal{A}^\alpha$, $\cal{B}^\alpha$ &
$\hat{O}^\alpha(\vec{r})$ &
$\hat{O}^\alpha_{ljl'j'}(r)$ &
$\kappa_{\alpha\beta}$\\
\hline
&&&&\\
$\delta\rho_q$ &
$ \left(
\begin{array}{cc}
2 & 0 \\
0 & 0
\end{array} 
\right)$
,
$ \left(
\begin{array}{cc}
1 & 0 \\
0 & -1
\end{array} 
\right)$
 &
1 &
 1 &
$ a_{qq'}$ \\
&&&&\\
$\delta\Delta_+\rho_q$ & & 
$\larw{\Delta}+\rarw{\Delta}$ & 
$\ddell{r}+\ddelr{r}-\frac{l'(l'+1)}{r^2}-\frac{l(l+1)}{r^2}$ &
$b_{qq'}$ \\
&&&&\\
$\delta\vec{\nabla}\rho_q$ & & 
$\larw{\nabla}+\rarw{\nabla}$ & 
$\left\{
\begin{array}{l}
\sqrt{\frac{L}{2L+1}}\left(\dell{r}+\delr{r}+\frac{L-1}{r}\right)
\hspace{3pt}(\mbox{for}\hspace{3pt} {[\delta\vec{\nabla}\rho_q]^{\lambda=L-1}_L})
\\
-\sqrt{\frac{L+1}{2L+1}}\left(\dell{r}+\delr{r}-\frac{L+2}{r}\right)
\hspace{3pt}(\mbox{for}\hspace{3pt}{[\delta\vec{\nabla}\rho_q]^{\lambda=L+1}_L})
\end{array}
\right.$ & 
$c_{qq'}$ \\
&&&&\\
$\delta\tau_q$ & & 
$\larw{\nabla}\idot\rarw{\nabla}$ & 
$
\begin{array}{ll}
\dell{r}\delr{r}-\frac{1}{r}\delr{r}-\dell{r}\frac{1}{r}
\\
\hspace{0.5cm}
+\frac{l(l+1)+l'(l'+1)-(L+2)(L-1)}{2}\left(\frac{1}{r^2}\right)
\end{array}
$ & --
\\
&&&&\\
$\delta 2i\vec{j}_q$ 
& 
$ \left(
\begin{array}{cc}
2 & 0 \\
0 & 0
\end{array} 
\right)$
,
$ \left(
\begin{array}{cc}
1 & 0 \\
0 & 1
\end{array} 
\right)$
& 
$-\larw{\nabla}+\rarw{\nabla}$ & 
$\left\{
\begin{array}{l}
-\sqrt{\frac{L}{2L+1}}\left(\dell{r}-\delr{r}+\frac{l(l+1)-l'(l'+1)}{L}\frac{1}{r}\right)
\hspace{3pt}(\mbox{for}\hspace{3pt}{[\delta 2i\vec{j}_q]^{\lambda=L-1}_L})
\\
\sqrt{\frac{L+1}{2L+1}}\left(\dell{r}-\delr{r}-\frac{l(l+1)-l'(l'+1)}{L+1}\frac{1}{r}\right)
\hspace{3pt}(\mbox{for}\hspace{3pt}{[\delta 2i\vec{j}_q]^{\lambda=L+1}_L})
\end{array}
\right.$ &
$-b_{qq'}$ \\
&&&&\\
\hline
&&&&\\
$\delta\tilde{\rho}_{+q}$ & 
$ \left(
\begin{array}{cc}
0 & 1 \\
1 & 0
\end{array} 
\right)$
&
1 &
 1 &
$\tilde{a}_{q}\delta_{qq'}$ \\
 &&&&\\
$\delta\tilde{\rho}_{-q}$ & 
$ \left(
\begin{array}{cc}
0 & 1 \\
-1 & 0
\end{array} 
\right)$
 &
 1&
   1&
$-\tilde{a}_{q}\delta_{qq'}$ \\
\hline
\vspace{0.25cm}
\\
\end{tabular}
\caption{The correspondence and the expressions 
for the matrices $\mathcal{A}^\alpha$ and 
$\mathcal{B}^\beta$, 
the operators $\hat{O}^\alpha$ and
$\hat{O}^\alpha_{ljl'j'}$ and $\kappa_{\alpha\beta}$
appearing in Eqs.~\refp{eq:resintmat}, 
\refp{eq:resfunc} and \refp{eq:resfunc2}. See also the text and
Appendix.
\label{tb:optable}}
\end{table}
\end{center}
\end{widetext}

\section{\label{sec:calc}Numerical analysis}

In this section, we shall demonstrate the Skyrme continuum QRPA by performing
numerical calculations for the dipole and quadrupole 
responses in $^{20}$O and $^{54}$Ca.

\subsection{\label{sec:input} Numerical procedure}

Let us first describe the detailed procedure of the numerical
calculation.

We adopt the SkM$^*$ parameter set of the Skyrme interaction  and the mixed-type parametrization
of the DDDI pairing interaction 
($\eta=0.5$, $\gamma=1$, $\rho_0=0.16$ fm$^{-3}$)\cite{dobaczewski-pair}
 for most of the calculations. 
The force strength $V_0$ of the DDDI is chosen so that  the
average neutron pairing gap $\langle \Delta_n \rangle$ 
reproduces the overall magnitudes
of the experimental odd-even mass differences for the isotopic chain, obtained 
with the three-point formula\cite{satula}.
Here we use the average pairing gap defined by
$\bra \Delta_n\ket=\int d\vec{r} 
\tilde{\rho}_n(\vec{r})\Delta_n(\vec{r})/\int d\vec{r}\tilde{\rho}_n(\vec{r})$.
The adopted value is $V_0=-280$ and $-285$ MeVfm$^{-3}$ 
for $^{20}$O and $^{54}$Ca
producing  $\bra \Delta_n\ket=$ 1.91 MeV and 1.29 MeV, respectively.

Since we use the contact interaction for the effective pairing
interaction, we need a cut-off of the quasi-particle states
in the HFB calculation. We define the cut-off
with respect to the quasi-particle energy 
$E_\alpha < E_{max} = 60$ MeV.
Concerning the angular momentum quantum numbers $lj$ we
sum up the quasi-particle states up to 
$l_{max}=7\hbar$ and $8\hbar$ for $^{20}$O and $^{54}$Ca, 
respectively. 
In performing the HFB and the continuum QRPA calculations, we discretize
the radial coordinate space  up to $r_{max}=15$ fm with an equidistant 
interval
$\Delta r=0.2$ fm. 
In the continuum QRPA calculations, the dynamical quantities to
be obtained are the eighteen functions
$[\delta\rho_{q}]_L,[\delta\tilde{\rho}_{\pm q}]_L,
[\delta\Delta_+\rho_q]_L,[\delta\vec{\nabla}\rho_q]_L^{\lambda=L\pm1},
[\delta 2i\vec{j}_q]_L^{\lambda=L\pm1}$ and $[\delta\tau_q]_{L}$
which obey the linear response equation (\ref{eq:lineq2}).
Using the same radial mesh, these functions are represented as
a grand vector while the linear response equation is represented
as a linear algebraic equation where the response function
$\tilde{R}^{\alpha\beta}_{0,qL}$ (and $S_q^{\alpha\beta}$)
corresponds to a matrix.
Since the number of the functions to be solved 
is larger (18 vs. 6) than in the previous continuum QRPA 
that handles only the local densities
$[\delta\rho_q]_L$  and $[\delta\tilde{\rho}_{\pm q}]_L$, the number of
the matrix elements of the response
functions is therefore about ten times larger 
than in the previous continuum QRPA calculations.
To reduce the increased computational cost thus caused,
we have chosen 
the values of $l_{max}$ and $r_{max}$ smaller than those
used in our previous calculations
\cite{matsuo01,matsuo02,matsuo-mizu-seri,matsuo06}.
For the same reason we have used here
a relatively large smearing parameter $\epsilon=1.0$ MeV
 in most of the following calculations. 
We evaluate the strength function at discretized excitation
energies with an interval of 0.5MeV. 

It is noted here that the self-consistency is not completely satisfied in the
present formulation since a few approximations
are introduced in deriving the residual interaction from 
the Skyrme HFB functional.
Consequently the spurious modes of motion which 
should have exact zero excitation energy
according to the Thouless's theorem\cite{ringschuck} do not emerge 
at the expected energy. 
A commonly adopted procedure to circumvent this problem is to 
renormalize the residual interaction 
$\kappa_{\alpha\beta}$ in Eq.(\ref{eq:lineq2}) by an overall 
factor $f$ as 
$\kappa_{\alpha\beta} \rightarrow f\times\kappa_{\alpha\beta}$ so that
the excitation energy of the spurious mode is forced at 
the zero
energy\cite{shlomo,matsuo01,matsuo02,matsuo-mizu-seri,yamagami2,yamagami-aizu,yamagami3,khan,khan2}.
We apply this renormalization procedure to the particle-hole
residual interactions that are derived from 
the Skyrme HF functional $E_{Skyrme}$.
The residual interaction in the particle-particle channel,
derived from the pair correlation 
energy $E_{pair}$, is kept in the original strength
since the continuum QRPA in the Green's function formalism
fulfills the self-consistency 
in the particle-particle channel with high accuracy
\cite{matsuo01,matsuo02}.
The renormalization
factor is $f=1.0470$ and 1.0142 for  $^{20}$O and $^{54}$Ca, respectively.

In the following analysis, we would like to demonstrate how the description
of the multipole response is improved in comparison with the previous 
continuum QRPA where  the residual
interaction is simplified to a contact force.
For this purpose, we perform calculations where the Landau-Migdal (LM)
approximation to the residual interaction is introduced
\cite{khan,khan2,khan3,khan-goriely,yamagami-aizu,
yamagami2,yamagami3,matsuo06}. 
This is an approximation which replaces the residual interaction
by a contact force $\propto \delta(\vec{r}-\vec{r}')$ whose strength
is given by the density-dependent 
Landau-Migdal parameters $F_0$ and $F_0'$ evaluated for
the Skyrme  functional\cite{LM-Skyrme,GiaiSagawa,bender}  
using the local density approximation. 
It should be noted, however, that the Landau-Migdal
parameters $F_0$ and $F_0'$ contain a part of the $t_1$ and $t_2$
terms, and this approximation should be distinguished from
dropping all the $t_1$ and $t_2$ terms.

\subsection{\label{sec:strength}Strength function}
\begin{figure}
\includegraphics[scale=0.65,angle=270]{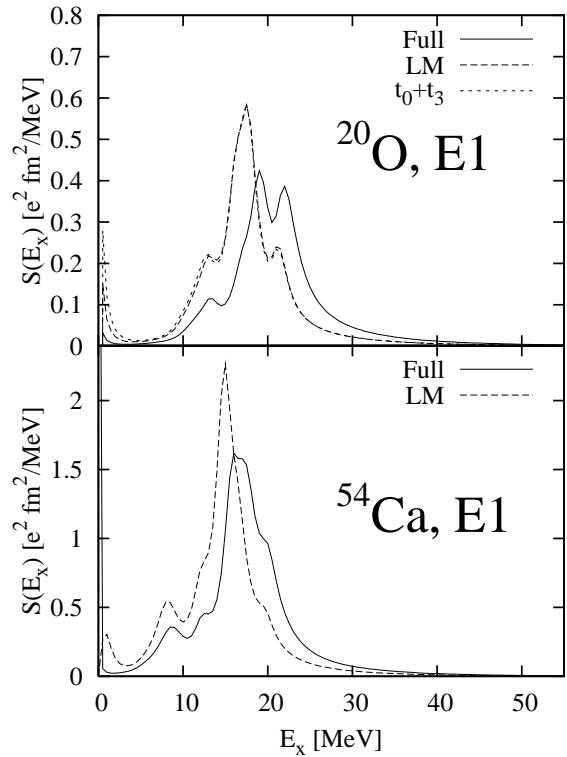}
\caption{\label{fig:e1st-o20ca54} The $B(E1)$ strength function 
of isovector dipole response in $^{20}$O (upper panel) and in
$^{54}$Ca (lower panel) calculated with the parameter set SkM$^*$. 
The solid curve is the result obtained in the full calculation while the 
dashed curve is that in the Landau-Migdal (LM) approximation. 
The dotted curve in the upper panel is the one in the $t_0+t_3$
 approximation. See also the text.
}
\end{figure}
\begin{figure}
\includegraphics[scale=0.6,angle=270]{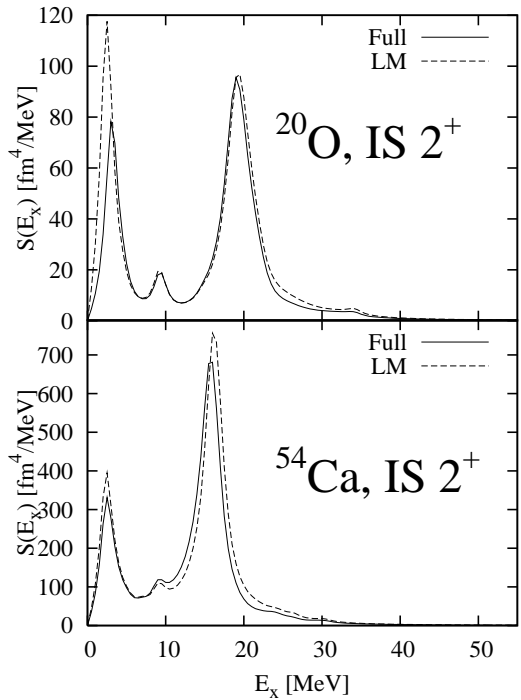}
\caption{\label{fig:L2st-is-o20ca54} The same as
 Fig.\ref{fig:e1st-o20ca54} 
but for the $B$(IS2) isoscalar quadrupole strength function
in $^{20}$O and $^{54}$Ca.}
\end{figure}
\begin{figure}
\includegraphics[scale=0.5,angle=270]{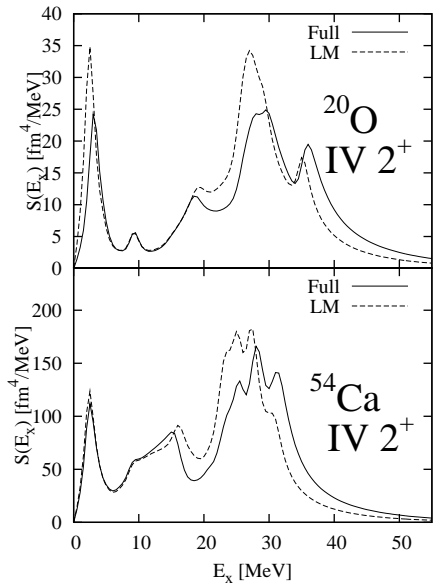}
\caption{\label{fig:L2st-iv-o20ca54} The same as
 Fig.\ref{fig:e1st-o20ca54} 
but for the $B$(IV2) isovector quadrupole strength function.}
\end{figure}

The strength function 
\begin{eqnarray}
S(\hbar\omega)
&\equiv&\sum_{\nu,M} |\bra \nu|\hat{F}_{LM}|0\ket|^2\delta(\hbar\omega-E_\nu)
\nonumber
\\
&=&-\frac{2L+1}{\pi}\mbox{Im}\sum_q\int dr f_{qL}(r)
[\delta\rho_{q}]_L(r,\omega)
\end{eqnarray}
for the operator $\hat{F}_{LM}$ with the multipolarity
$L$ 
can be evaluated in terms of the solution 
$[\delta\rho_q]_L(r,\omega)$ of the linear response equation
(\ref{eq:lineq2}) obtained for the external field $\hat{F}_{L0}$. 
We evaluate the $B$(E1), $B$(IS2) and $B$(IV2)  strength functions
associated with the electric dipole operator
\begin{eqnarray}
  \hat{F}^{IV}_{1M}
  =\frac{eN}{A}\sum_{i=1}^Z r_i Y_{1M}(\Omega_i)-\frac{eZ}{A}\sum_{i=1}^N r_i Y_{1M}(\Omega_i)
\end{eqnarray}
and the isoscalar/isovector quadrupole operators
\begin{eqnarray}
  \hat{F}^{IS}_{2M}
  =\sum_{i=1}^A r_i^2 Y_{2M}(\Omega_i),
\hspace{0.2cm}
  \hat{F}^{IV}_{2M}
  =\sum_{i=1}^A \tau_z r_i^2 Y_{2M}(\Omega_i).
\end{eqnarray}

The $B$(E1) strength functions calculated for $^{20}$O and $^{54}$Ca are shown 
with the solid curve in Fig.\ref{fig:e1st-o20ca54}. 
 The broad peaks around $E_x=20$ MeV in $^{20}$O and $E_x=16$ MeV in $^{54}$Ca
 correspond to the giant dipole resonance (GDR). 
There is s small bump around
 $E_x=8$ MeV in $^{54}$Ca, which corresponds to 
the soft dipole excitation or the pygmy dipole resonance.
We find however that the small peak $E_x=13$ MeV in $^{20}$O is
neither the GDR nor the soft dipole excitation,
but rather a non-collective two quasiparticle excitation
(cf. Section \ref{sec:trdnst}).
The strength at $E \approx 0$ is due to the spurious mode, and it 
is caused by the incomplete self-consistency.

For the sake of comparison, the $B$(E1) strength function 
obtained in the Landau-Migdal (LM) 
approximation of the residual interaction is also
plotted with the dashed curve in Fig.\ref{fig:e1st-o20ca54}.
Note that the renormalization factor used in
the Landau-Migdal approximation 
($f=0.6686$ and 0.7515 for $^{20}$O and $^{54}$Ca, respectively)
deviate significantly from one in contrast to those
in the full calculation ($f=1.0470$ and 1.0142). 
This fact suggests that there is significant
improvement in the self-consistency 
compared with the LM approximation. This feature
is pointed out in a Skyrme-QRPA calculation using
discretized continuum quasiparticle states\cite{yamagami-aizu}.

It is seen in Fig.\ref{fig:e1st-o20ca54} that
 the profile of the strength function obtained in the LM approximation
differs significantly from that in the full calculation.
The peak positions of the giant dipole resonance
are apparently different.
Estimating the centroid energy of the GDR
by $E({\rm GDR})=m_1/m_0$ using the energy weighted sum $m_1$ 
and the non-weighted
 sum $m_0$, we find $E({\rm GDR})=20.66$ MeV ($^{20}$O) and 15.80 MeV 
($^{54}$Ca) for the full calculation
while $E({\rm GDR})=17.67$ and 14.20 MeV in the LM approximation, exhibiting
a rather large difference by about 2-3 MeV. 
It is clear that the LM approximation is not very appropriate 
to give precise quantitative
description of the GDR.

If we evaluate the energy weighted sum integrated up to $E=15$ MeV
for $^{20}$O,
the full calculation 
gives 9.7 \%  of the classical Thomas-Reiche-Kuhn 
(TRK) sum rule value while
it is 20.1\% in the LM approximation. Comparing with the 
experimental value  12\%\cite{gsi-oxygen}, we find that the
full calculation is in better agreement with the experiment.
The $B$(E1) strength function in $^{20}$O 
is calculated in a fully
self-consistent Skyrme-QRPA calculation\cite{terasaki2} using the same 
SkM$^*$. 
We find only small difference between our calculation
and that in Ref.\cite{terasaki2}. It may be attributed to the
neglect of the Coulomb and spin-dependent terms in our calculations.
The observed effect of the velocity dependent terms 
on the GDR centroid energy is essentially the same as that 
discussed in Ref.\cite{yamagami-aizu}.

Figure \ref{fig:L2st-is-o20ca54} displays the  $B$(IS2) isoscalar strength
function for the quadrupole responses in $^{20}$O and $^{54}$Ca.
In both nuclei there are two significant peaks, one around $E_x= 2-3$ MeV
corresponding to the low-lying $2^+$ collective vibrational mode and the 
other around $E_x =15-20$ MeV 
corresponding to the isoscalar giant quadrupole resonance (ISGQR). 
(The experimental  $2_1^+$ energy in $^{20}$O is
1670 keV\cite{Thirolf}.) 
  The calculated isoscalar quadrupole strength 
function for $^{54}$Ca is quite similar to that
obtained in the fully self-consistent Skyrme QRPA 
using the same SkM$^*$\cite{terasaki2} 
apart from features associated with different
choices of the smoothing width.

Concerning the effect of the velocity dependent terms,
it is seen in Fig.~\ref{fig:L2st-is-o20ca54} 
that the difference between the full calculation and the
LM approximation is less significant in
comparison with the isovector dipole response: 
the peak positions of the giant isoscalar quadrupole 
resonance ($E=19.0$ MeV) and of the low-lying state
($E= 3.0$ MeV) in $^{20}$O is affected only little by 
inclusion of the velocity dependent terms. 
The same is seen also in $^{54}$Ca.
Note however that 
the influence of the velocity dependent terms on the
$B$(IV2) isovector distribution is clearly
 larger than in the case of the $B$(IS2) isoscalar  strength
distribution as is seen in Fig.\ref{fig:L2st-iv-o20ca54}.
Combining Figs. \ref{fig:e1st-o20ca54}, \ref{fig:L2st-is-o20ca54} and 
\ref{fig:L2st-iv-o20ca54}, we see an apparent trend that 
the influence of the velocity-dependent terms is more significant in
the isovector responses
than in the isoscalar responses.

Before moving to the next subsection, we would like to make a
few additional remarks on the effect of the velocity dependent terms.
We first remark that
it is possible to consider another way to evaluate the effect of the
velocity dependent terms, e.g. by comparing with a calculation 
where all the velocity dependent
terms containing the $t_1$ and $t_2$ parameters are completely
neglected. (In other words, it is the calculation where only
the simple contact interaction associated with the
$t_0$ and $t_3$ terms are taken into account. 
It is different from the Landau-Migdal (LM) approximation
since in the latter a part of the $t_1$ and $t_2$ terms is 
renormalized into the Landau-Migdal parameters $F_0$ and $F_0'$.)
The $B$(E1) strength function in this $t_0+t_3$ approximation
calculated for $^{20}$O is plotted in the upper panel of
Fig.\ref{fig:e1st-o20ca54} 
together with the other two curves representing the full calculation
and the LM approximation. We find here that the result obtained 
in the $t_0+t_3$ approximation is almost identical to that 
in the LM approximation. 
Secondly, we remark that the effect of the velocity dependent
terms depends on the adopted Skyrme parameter set. To demonstrate
this we show in Fig.4 the $B$(E1) strength function in $^{20}$O
obtained with SLy4\cite{SLy4} instead of SkM$^*$. It is seen that
there is no big difference in the GDR peak position
between the full calculation and in the Landau-Migdal approximation,
and hence the effect of the velocity dependent terms
in the case of SLy4 appears smaller than in the case of SkM*.
Note however that even in this case there is significant difference
between the LM and $t_0+t_3$ approximations. If 
we look at the difference between the full calculation and the 
$t_0+t_3$ approximation, the effect of the velocity dependent terms
is not negligible. Note also that the difference between the full 
calculation and the LM approximation is not negligible in the sum rule 
(cf. next subsection). 

\begin{figure}
\includegraphics[scale=0.65,angle=270]{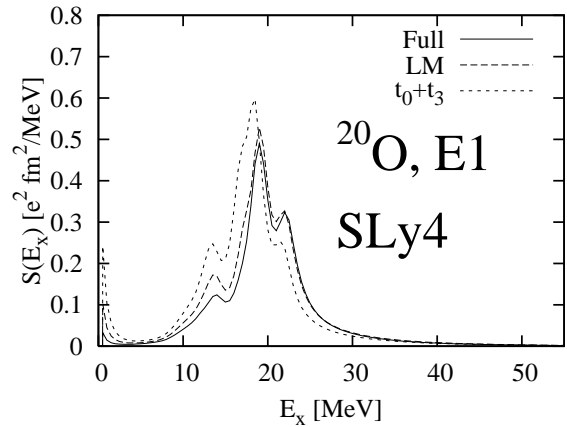}
\caption{\label{fig:E1st-o20-SLy4}
The same as the upper panel of Fig.\ref{fig:e1st-o20ca54}, 
but the Skyrme parameter set SLy4 is used.
}
\end{figure}

\subsection{\label{sec:sumrule}Energy weighted sum rule}

Let us  analyze whether 
the energy weighted sum rule is satisfied in the present
calculations. For this purpose
we  evaluate the running energy weighted sum defined by
\begin{eqnarray}
W(E_x)=\int_0^{E_x} \hspace{-0.3cm}dE \hspace{0.1cm}E \hspace{0.1cm}S(E),
\end{eqnarray}
which integrates the sum up to an excitation energy $E_x$.
The limiting value of $W(E_x)$ for a sufficiently large $E_x$ is to be
compared with the energy-weighted sum rule (EWSR).

The EWSR for the 
$B$(IS$L$) isoscalar multipole 
strength function  is 
identical to the
classical sum rule
$m_1^{cl}\equiv \frac{1}{4\pi}\frac{\hbar^2}{2m}
 L(2L+1)^2 A \bra r^{2L-2} \ket$ which is expressed in terms of
the expectation value of the radial moment $r^{2L-2}$ with respect
to the ground state\cite{BM2,ringschuck}.
For the isovector multipole strength functions, however, 
the EWSR contains the enhancement factor which arises from the residual interaction,
in particular the velocity-dependent terms in the case of the Skyrme 
effective force\cite{tsai,BM2,stringari,speth,vanderWourde}. 
The EWSR for the $B$(E1)
electric  dipole strength function
is given by 
$m_1^{\rm EWSR}({\rm E1})=\frac{NZ}{A^2}m_1^{cl}\left(1+\kappa\right)$
where $\kappa$ is the enhancement factor which is easily evaluated in the
case of the Skyrme force\cite{terasaki,shlomo2}. The value of $\kappa$
for SkM$^*$ is $\kappa=0.32$ and 0.36 in $^{20}$O and  $^{54}$Ca, respectively.

\begin{figure}
\includegraphics[scale=0.6,angle=270]{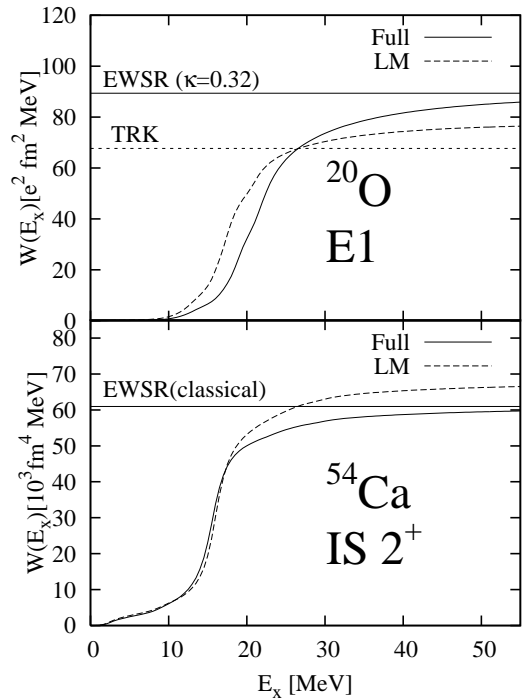}
\caption{\label{fig:esumrule} The running energy weighted
 sum for the $B$(E1) 
electric dipole strength function in $^{20}$O(upper panel) and 
that for the $B$(IS2) isoscalar 
quadrupole strength function 
in $^{54}$Ca(lower panel) obtained using the Skyrme parameter set SkM$^*$. The solid curve represents 
the result of the full calculation of the present Skyrme continuum QRPA
while the dashed curve is the result obtained in the Landau-Migdal
(LM) approximation to the residual interaction. 
The horizontal solid line indicates the
 value of the energy-weighted sum rule in the present calculation, which includes the enhancement
 factor in the isovector response. The dotted horizontal line in the upper
 panel denotes the value of the 
Thomas-Reiche-Kuhn sum rule.}
\end{figure}

The upper panel of the Fig.\ref{fig:esumrule} displays the running
energy 
weighted sum $W(E_x)$ for the
$B$(E1) strength function of the dipole response 
in $^{20}$O (cf. Fig.\ref{fig:e1st-o20ca54}).
The running sum evaluated at the highest calculated energy $E_x=55$ MeV
reaches 96\% of the EWSR. This suggests that 
the EWSR is 
satisfied in the present calculation.
It is noted
that $W(E_x)$ approaches 
the EWSR value including the enhancement factor,
but not the classical TRK value 
plotted with the dotted horizontal line in Fig.~\ref{fig:esumrule}.
Namely the 
effect of the velocity dependent terms in the residual interaction is 
indeed included in the present calculation.
In the same figure, 
we also show the result obtained in the LM approximation.
In this case, however, the running sum reaches only 86 \% of the
EWSR, and hence the approximation fails to describe the EWSR and the enhancement factor.
The fluctuations $\delta\Delta_+\rho$, $\delta\vec{\nabla}\rho$,
$\delta\tau$ and, in particular, $\delta 2i \vec{j}$ play 
the essential role to restore
the EWSR since these are the 
fluctuations associated with the velocity-dependent 
terms ($\propto t_1$ and $t_2$).
Note that the LM approximation neglects these fluctuations
although a part of the velocity-dependent $t_1 + t_2$ terms is
taken into account via the Landau-Migdal parameters $F_0$ and $F'_0$.

The lower panel in Fig.\ref{fig:esumrule} displays 
the running energy weighted sum for the $B$(IS2) isoscalar quadrupole
strength function in $^{54}$Ca (cf. Fig.\ref{fig:L2st-is-o20ca54}). 
The energy weighted sum
amounts to 98\% of the EWSR, and we confirm more clearly than 
in the analysis of 
the electric dipole strength that 
the EWSR is satisfied in the present calculation. 
When we adopt the LM
approximation where the fluctuations
$\delta\Delta_+\rho$, $\delta\vec{\nabla}\rho$, $\delta\tau$
and $\delta 2i \vec{j}$ are neglected, 
the running sum $W(E_x)$ at the maximum energy
$E_x=55$ MeV overshoots the EWSR  by about ten percent. Again
the consistent inclusion of 
the velocity dependent terms 
in the QRPA description is essential to guarantee the EWSR.
 It is noted also that the difference between
the full calculation and the Landau-Migdal approximation is significant
mainly in the energy region 
($E_x > 18$ MeV) higher than the giant resonance peak.

 We confirmed that the EWSR is satisfied 
also in the case of the other Skyrme parameter set SLy4.
For the $B$(E1) and $B$(IS2) strength functions in $^{20}$O,
the running sum W($E_x$) at the highest energy $E_x=55$ MeV
reaches 95\% and 97\% of the EWSR, respectively. In the
LM approximation, on the contrary, the sum overshoots
the EWSR by 6\% and 12\% in the $B$(E1) and $B$(IS2) strength
functions, respectively.

\begin{figure}
\includegraphics[scale=0.6,angle=270]{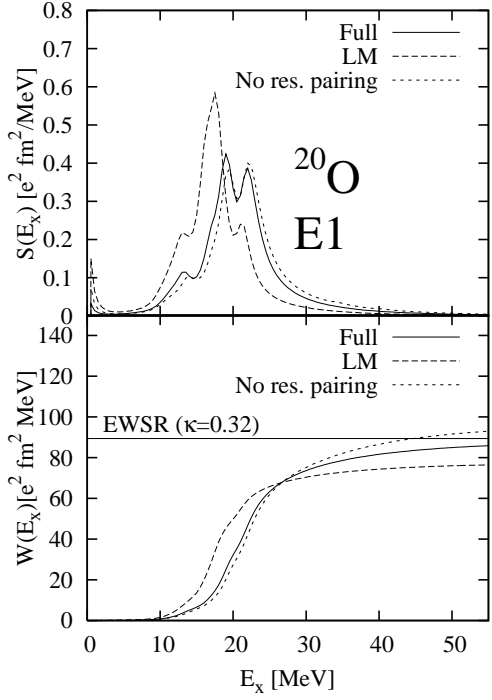}
\caption{\label{fig:e1st-o20-nodp} The $B(\mbox{E1})$ strength function(upper panel) and
 its running energy-weighted sum(lower panel) for the isovector
 dipole response of $^{20}$O using SkM$^*$. 
The solid and dashed curves are the results obtained in the full
 calculation and in the Landau-Migdal approximation, respectively, while
 the dotted curve is the one in which we neglect the residual pairing
 interaction in the continuum QRPA.
}
\end{figure}

In Fig.\ref{fig:e1st-o20-nodp}, we demonstrate influence  of the residual
pairing interaction on the energy-weighted sum. When the
residual pairing interaction is dropped in the continuum QRPA
calculation (i.e., the dynamical pairing
effect is neglected), 
the energy-weighted sum overestimates the EWSR value by 5\%.  
As pointed out already\cite{matsuo01,matsuo02,paar},
inclusion of the dynamical pairing effect is important
to guarantee the energy weighted sum-rule because otherwise the 
self-consistency in the pairing channel would be violated.
This means, in the present context, that we need to include
both the velocity dependent terms of the Skyrme effective interaction and the
residual pairing interaction.

\begin{figure}
\includegraphics[scale=0.6,angle=270]{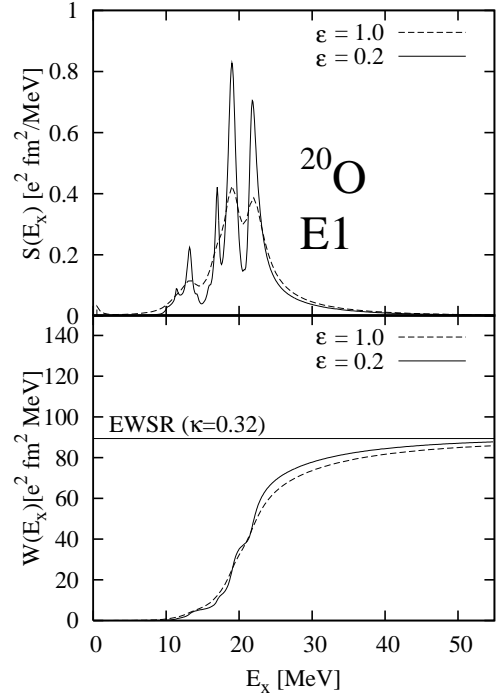}
\caption{\label{fig:e1st.gm02.o20} 
The $B$(E1) strength function in $^{20}$O (upper panel) and its
running energy weighted sum (lower panel) which are obtained 
with a small smoothing width
parameter $\epsilon$=0.2 MeV. For comparison, the results obtained with
$\epsilon$=1.0 MeV, shown already in Fig.\ref{fig:e1st-o20-nodp}, are
also plotted with the dashed curves.  
}
\end{figure}

\begin{figure}
\includegraphics[scale=0.6,angle=270]{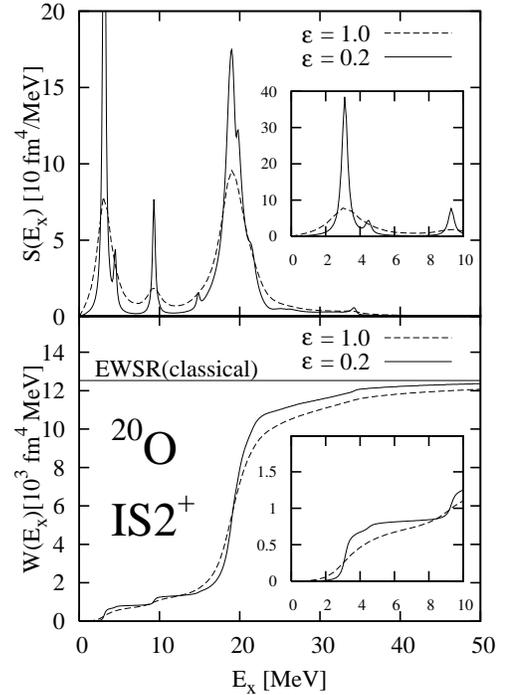}
\caption{\label{fig:is2st.gm02.o20} 
The same as Fig.\ref{fig:e1st.gm02.o20}, but for the
$B$(IS2) strength function in $^{20}$O.  
}
\end{figure}

Figures \ref{fig:e1st.gm02.o20} and \ref{fig:is2st.gm02.o20} show the
$B({\rm E1})$ and $B({\rm IS2})$ strength functions in $^{20}$O, respectively,
calculated with use of a small smearing width $\epsilon=0.2$ MeV.
They are compared with those obtained with $\epsilon=1.0$ MeV (cf. 
Figs.\ref{fig:e1st-o20ca54} and \ref{fig:L2st-is-o20ca54}). The
running energy weighted sum of the strength function is also shown.
Finer structures of the strength functions are visible here: we can
distinguish the first excited $2^+$, which is a bound discrete state
located below the separation energy. 

It is seen from the bottom panels of Figs. \ref{fig:e1st.gm02.o20} and \ref{fig:is2st.gm02.o20} 
that the agreement with the energy weighted sum rule is improved
with use of the smaller smearing width. The running sums at the
highest energy $E_x=55(50)$ MeV are 98 and 98 \%
for the $B({\rm E1})$ and $B({\rm IS2})$ strengths, respectively,
which should be compared with the corresponding values 96 \% and 96 \% obtained with
$\epsilon=1.0$ MeV. 

In the following we return to $\epsilon=1.0$ MeV since
calculations with $\epsilon=0.2$ MeV demand very long computation time.

\subsection{\label{sec:trdnst}Transition densities}

Let us now look into individual modes of excitation. For this purpose we
analyze the transition densities associated with each excitation mode. 
We evaluate three kinds of
transition densities\cite{matsuo-mizu-seri,matsuo-cooper,khan,khan2} 
\begin{eqnarray}
\rho^{ph}_{iq}(\vec{r})&=&
\left\bra\Phi_i\right|\sum_\sigma\psi^\dag_q(\vec{r}\sigma)
\psi_q(\vec{r}\sigma)\left|\Phi_0\right\ket=Y_{LM}^*(\hat{\vec{r}})\rho^{ph}_{iqL}(r),\nonumber\\
\\
P^{pp}_{iq}(\vec{r})&=&\left\bra\Phi_i\right|\psi_q^\dag(\vec{r}\uparrow)
\psi_q^\dag(\vec{r}\downarrow)\left|\Phi_0\right\ket=Y_{LM}^*(\hat{\vec{r}})
P^{pp}_{iqL}(r), \nonumber\\
\\
P^{hh}_{iq}(\vec{r})&=&\left\bra\Phi_i\right|\psi_q(\vec{r}\downarrow)
\psi_q(\vec{r}\uparrow)\left|\Phi_0\right\ket=Y_{LM}^*(\hat{\vec{r}})
P^{hh}_{iqL}(r)\nonumber\\
\end{eqnarray}
using the solution of the linear 
response equation (\ref{eq:lineq2}) at an energy corresponding to a
peak in the strength function.
Here $\rho^{ph}_{iq}(\vec{r})$ is the
usual particle-hole transition density while 
$P^{pp}_{iq}(\vec{r})$ and $P^{hh}_{iq}(\vec{r})$ are the particle-pair
and hole-pair transition densities associated
with the pair addition and removal amplitudes, respectively.

\begin{figure}
\includegraphics[scale=0.52]{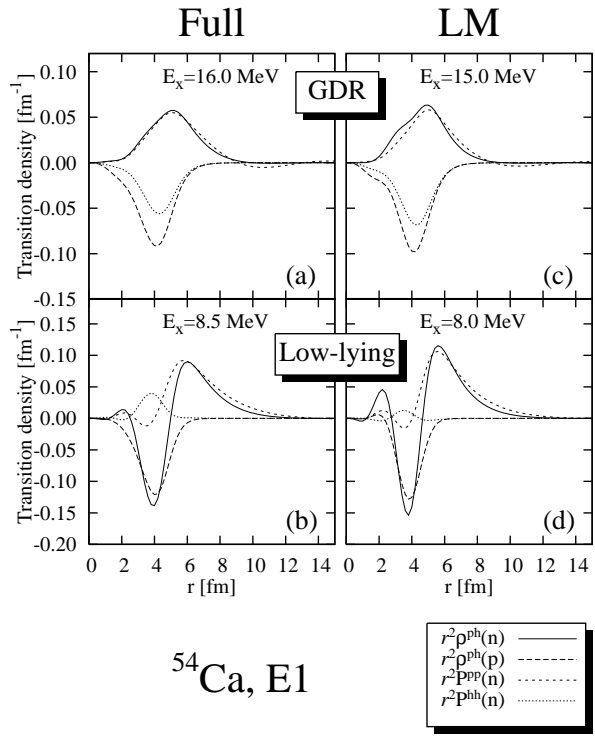}
\caption{\label{fig:e1trd-ca54} The transition densities 
$r^2\rho^{ph}_{iqL}(r), r^2 P^{pp}_{iqL}(r)$ and $r^2 P^{hh}_{iqL}(r)$ 
for the isovector dipole response
in $^{54}$Ca. 
(a) and (b) are those in the full calculation, evaluated at 
$E_x=16.0$ MeV (the peak energy of the GDR) and
$E_x=8.5$ MeV (the peak energy of the soft dipole excitation), respectively.
(c) and (d) are those in the Landau-Migdal approximation.
We here plot the neutron amplitudes weighted with the volume element
$r^2$ for the three kinds of
transition densities. For the proton we plot only
$r^2\rho^{ph}_{iqL}(r)$. 
Note that we need to normalize the transition densities 
in terms of the $B$(E1) strength of this mode\cite{matsuo-mizu-seri}, but 
we assume here the unit strength $B$(E1)=1 e$^2$fm$^2$.}
\end{figure}

Let us first discuss the giant dipole resonance (GDR) and the low-lying 
small peak seen in the $B$(E1) strength function. We 
focus on $^{54}$Ca,
where we find two peaks at $E_x=8.5$ MeV and 16.0 MeV 
in the strength function shown in Fig.\ref{fig:e1st-o20ca54}.
(The corresponding
peaks in the Landau-Migdal approximation are found at $E_x=8.0$ and
15.0 MeV.) The transition densities evaluated at $E_x=16.0$ MeV
is shown in Fig.\ref{fig:e1trd-ca54}(a). It is seen that the
particle-hole transition density $\rho^{ph}_{iqL}(r)$ is large 
around the nuclear surface, and that the neutron and
proton amplitudes have the opposite phases. 
This is indeed the feature typical of the isovector
giant dipole resonance (GDR). The character of
the excitation mode  at $E_x=8.5$ MeV is 
different from that of the giant resonance. This is seen in
the transition densities 
shown in Fig.\ref{fig:e1trd-ca54}(b), where we find a 
characteristic feature
 that the neutron amplitude 
of the particle-hole transition density dominates over the proton's 
in the exterior of the nucleus. The neutron
amplitude has a node around the nuclear surface, and it exhibits 
significant magnitude also inside the surface, where the proton 
amplitude also has
 comparable magnitude with the same phase as that of the neutron.
This is the feature which is often interpreted as the soft 
dipole excitation or the pygmy dipole resonance characteristic
to neutron-rich nuclei\cite{catara,hamamoto2,Vretenar2,matsuo-mizu-seri,paar}.

It is interesting to check
how well the Landau-Migdal approximation can describe 
the transition densities and the mode characters.
This is a non-trivial question since we have already seen
that there is rather large difference in the $B$(E1) strength functions
between in the full calculation and the Landau-Migdal approximation. 
The transition densities evaluated at 
$E_x=$15.0 and 8.0  MeV are shown in the 
panels (c) and (d) in Fig.\ref{fig:e1trd-ca54}, which are
compared with those in (a) and (b). 
In both the cases of the giant dipole resonance ((a) vs. (c))
and of the soft dipole excitation ((b) vs. (d)), 
the basic features of the transition 
densities are the same in the full and LM calculations, although we see
small but non-negligible  differences, e.g.,
in the relative
size between the particle-hole transition density $\rho^{ph}(r)$ and
the particle-pair transition density $P^{pp}(r)$ for the neutrons. 
This comparison suggests that the Landau-Migdal approximation can
be used to describe the basic structure of these excitation modes while
some reservation should be held when aiming at a quantitative description.

\begin{figure}
\includegraphics[scale=0.52]{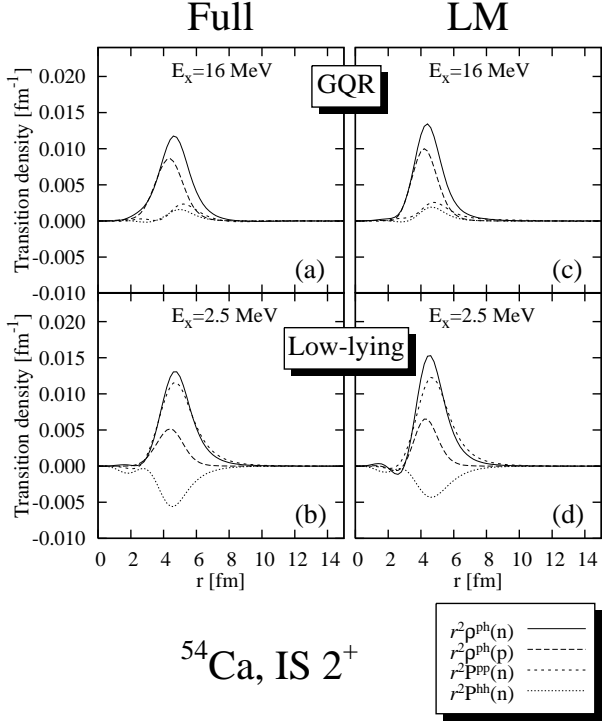}
\caption{\label{fig:L2trd-ca54} The same as 
Fig.\ref{fig:e1trd-ca54} but for the isoscalar quadrupole
response in $^{54}$Ca. 
(a) and (b) are those in the full calculation, evaluated at 
$E_x=16.0$ MeV (the peak energy of the ISGQR) and
$E_x=2.5$ MeV (the peak energy of the low-lying surface vibration), 
respectively.
(c) and (d) are those in the Landau-Migdal approximation.
Note that we here normalized the transition densities 
in terms of the unit strength $B$(IS2)=1 fm$^4$.
}
\end{figure}

Similar analysis of the transition densities is 
performed also for the isoscalar quadrupole modes in $^{54}$Ca.
We analyze here 
the isoscalar giant quadrupole resonance having a broad peak around
$E_x=16.0$ MeV and the 
low-lying quadrupole vibrational state peaked around $E_x=2.5$ MeV.
The transition densities evaluated at these peak energies are plotted
in Fig.\ref{fig:L2trd-ca54} (a) and (b). The transition densities of 
the corresponding peaks
in the Landau-Migdal approximation are also plotted in (c) and (d).
Comparing  the full calculation and the Landau-Migdal approximation, 
we observe the same trends as found 
in the case of the giant dipole resonance and the
soft dipole excitation. 
Namely there is no difference in the basic features of the modes, 
but quantitative details
of the transition densities depend on whether the velocity-dependent
terms of the Skyrme interaction are taken account or not. 
   

We analyzed also the transition densities evaluated at the
peaks of the dipole and quadrupole strength functions in $^{20}$O. 
We obtained results similar to those in $^{54}$Ca except 
for the low-lying peak at $E_x=8.5$ MeV in the dipole response. 
The transition 
densities 
corresponding to this peak is different from those of the GDR nor
those of the soft dipole excitation. We infer this peak 
as having non-collective nature. 

\section{Conclusions}

We have developed the continuum QRPA which is based 
on the Skyrme-Hartree-Fock-Bogoliubov energy functional.
In deriving the residual interaction used in the QRPA,
we have taken into account   the velocity-dependent central terms
(proportional to the $t_1$ and $t_2$ coefficients)
of the Skyrme effective interaction in order to guarantee
the energy weighted sum rule for the multipole responses,
but we neglected the two-body spin-orbit, the spin-spin and the Coulomb
interactions.

The new continuum QRPA is applied to the isovector dipole and
the isoscalar/isovector quadrupole responses of medium-mass
neutron rich nuclides $^{20}$O and $^{54}$Ca
using the SkM$^*$ parameter set.
It is confirmed numerically that the energy weighted sum rule is 
satisfied up to the enhancement factor relevant for the isovector 
responses. This
is because the velocity dependent terms are taken into
account in a way consistent with the Hartree-Fock-Bogoliubov
mean-fields of the ground state. 
We thus constructed the first Skyrme continuum QRPA formalism 
that satisfies the sum rule.

We have also examined the importance of the velocity dependent 
terms by comparing with the reduced calculation where 
the Landau-Migdal approximation is introduced to the residual interaction.
The continuum QRPA using the Landau-Migdal approximation 
gives an overall correct description
of the multipole responses, but influences of the approximation 
are seen in the shift of the centroid energy by 2-3 MeV 
of the isovector giant resonances and also in the violation of the
energy weighted sum rule by about 10-15 \% in the case of SkM$^*$.
We found also small but non-negligible influence 
in the transition densities for the low-lying dipole and quadrupole
modes. Note however that the qualitative features 
of the transition densities are described well even in this case, and
only very little influence is seen in the case of the giant resonances. 
The analysis gives a partial justification for the use of the 
Landau-Migdal approximation.

\begin{acknowledgments}
The authors thank T. Inakura for valuable discussions and giving us data
 of the strength function in RPA for debugging our numerical code. The
 numerical calculations were performed on the NEC SX-8 supercomputer
 systems at Yukawa Institute for Theoretical Physics, Kyoto University, and at Research
 Center for Nuclear Physics, Osaka University. The work was supported by
the Grant-in-Aid for Scientific Research(No.17540244) from the Japan
 Society for the Promotion of Science,   and also by 
the JSPS Core-to-Core Program, International
Research Network for Exotic Femto Systems(EFES).  

\end{acknowledgments}

\appendix

\section*{Details of the linear response equation\label{sec:appendix}}

Here we give some details of the
linear response equation (\ref{eq:lineq2}). 
The functions $a_{qq'},b_{qq'},c_{qq'}$ and $\tilde{a}_q$
appearing in Table \ref{tb:optable} and Eqs.(\ref{eq:resinthf})
and (\ref{eq:resintpair}) are expressed as follows in terms
of the effective interaction parameters and the local densities:
\begin{eqnarray*}
a_{qq'}
&=&
\left\{
\begin{array}{l}
\frac{1}{2}t_0(1-x_0)
\\
+
(\alpha+2)(\alpha+1)\frac{1}{12}t_3(1+\frac{1}{2}x_3)\rho^\alpha
\\
-
\frac{1}{12}t_3(x_3+\frac{1}{2})
\\
\times
\Big[
\alpha(\alpha-1)
\rho^{\alpha-2}\sum_{q''}\rho_{q''}^2
\\
\hspace{2.5cm}
+4\alpha\rho^{\alpha-1}\rho_q
+2\rho^\alpha
\Big]
\hspace{0.3cm}
(q=q')
\\
t_0(1+\frac{1}{2}x_0)
\\
+
(\alpha+2)(\alpha+1)\frac{1}{12}t_3(1+\frac{1}{2}x_3)\rho^\alpha
\\
-
\frac{1}{12}t_3(x_3+\frac{1}{2})
\\
\times
\left(
\alpha(\alpha-1)
\rho^{\alpha-2}\sum_{q''}\rho_{q''}^2
+2\alpha\rho^\alpha
\right)
\hspace{0.3cm}
(q\neq q')
\end{array}
\right.
\\
\\
b_{qq'}
&=&
\left\{
\begin{array}{c}
-\frac{1}{16}\left(t_1(1-x_1)+3t_2(1+x_2)\right) 
\hspace{0.5cm}(q=q')
\\
\\
-\frac{1}{8}\{t_1(1+\frac{1}{2}x_1)+t_2(1+\frac{1}{2}x_2)\} 
\hspace{0.5cm}
(q\neq q')
\end{array}
\right.
\\
\\
c_{qq'}
&=&
\left\{
\begin{array}{c}
-\frac{1}{16}
\left(t_1(x_1-1)+9t_2(x_2+1)\right)
\hspace{0.5cm}
(q=q')
\\
\\
\frac{1}{8}\{t_1(1+\frac{1}{2}x_1)-3t_2(1+\frac{1}{2}x_2)\}
\hspace{0.5cm}
(q\neq q')
\end{array}
\right.
\\
\\
\tilde{a}_q
&=&
\frac{V_0}{2}\left[1-\eta\left(\frac{\rho(r)}{\rho_0}\right)^\gamma\right].
\end{eqnarray*}


\begin{widetext}
\begin{center}
\begin{table}
\begin{tabularx}{\linewidth}{cc}
\hline
$\delta\rho_\alpha$ & 
$2\sum_{\beta\gamma}S^{\alpha\beta}_q
\left[
\tilde{\kappa}_{\beta\gamma}\delta\rho_{\gamma L}/r^2
+\delta_{\beta 0}f_{q L}
\right]
$ \\
\hline
&\\
$[\delta\rho_q]_L$ & $2\times\frac{2m_q^*}{\hbar^2}\rho_q \sum_{q'}b_{qq'}[\delta\rho_{q'}]_{L}$ \\
&\\
$[\delta\tilde{\rho}_{+q}]_L$ & $2\times\frac{2m_q^*}{\hbar^2}\tilde{\rho}_q\sum_{q'}b_{qq'}[\delta\rho_{q'}]_{L}$ \\
&\\
$[\delta\tilde{\rho}_{-q}]_L$ & 0 \\
&\\
$
\begin{array}{c}
[\delta\Delta_+\rho_q]_L
\\
\vspace{1.5cm}
\end{array}
$
& 
$
\begin{array}{l}
2\times\frac{2m_q^*}{\hbar^2}
\left\{
\rho_q\sum_{q'}a_{qq'}[\delta\rho_{q'}]_L+\tilde{\rho}_q\tilde{a}_q[\delta\tilde{\rho}_{+,q}]_L
+
2(\Delta\rho_q-2\tau_q)\sum_{q'}b_{qq'}[\delta\rho_{q'}]_{L} \right.
\\
+\rho_q\sum_{q'}(b_{qq'}-c_{qq'})([\delta\Delta_+\rho_{q'}]_L+2[\delta\tau_{q'}]_L)
+
\del{\rho_q}{r} \sum_{q'}b_{qq'}
\left(\sqrt{\frac{L}{2L+1}}[\delta\vec{\nabla}\rho_{q'}]_L^{\lambda=L-1}
-\sqrt{\frac{L+1}{2L+1}}[\delta\vec{\nabla}\rho_{q'}]_L^{\lambda=L+1}\right)
\\
\left.
-
\frac{2m_q^*}{\hbar^2}(\del{}{r}\frac{\hbar^2}{2m_q^*})\rho_q\sum_{q'}(b_{qq'}-c_{qq'})
\left(\sqrt{\frac{L}{2L+1}}[\delta\vec{\nabla}\rho_{q'}]_L^{\lambda=L-1}
-\sqrt{\frac{L+1}{2L+1}}[\delta\vec{\nabla}\rho_{q'}]_L^{\lambda=L+1}\right)
\right\}
+
2\times\frac{2m_q^*r^2}{\hbar^2}f_{qL}(r)
\end{array}
$
\\
&\\
$[\delta\vec{\nabla}\rho_q]_L^{\lambda=L-1}$ & 
$
2\times\frac{2m_q^*}{\hbar^2}\left\{
\sqrt{\frac{L}{2L+1}}\del{\rho_q}{r}\sum_{q'}b_{qq'}[\delta\rho_{q'}]_{L}
+
\rho_q\sum_{q'}(b_{qq'}-\frac{L}{2L+1}c_{qq'})[\delta\vec{\nabla}\rho_{q'}]_L^{\lambda=L-1}
+
\frac{\sqrt{L(L+1)}}{2L+1}\rho_q \sum_{q'}c_{qq'}[\delta\vec{\nabla}\rho_{q'}]_L^{\lambda=L+1}
\right\}
$
\\
&\\
$[\delta\vec{\nabla}\rho_q]_L^{\lambda=L+1}$ & 
$
2\times\frac{2m_q^*}{\hbar^2}\left\{
-\sqrt{\frac{L+1}{2L+1}}\del{\rho_q}{r}\sum_{q'}b_{qq'}[\delta\rho_{q'}]_{L}
+
\frac{\sqrt{L(L+1)}}{2L+1}\rho_q\sum_{q'}c_{qq'}[\delta\vec{\nabla}\rho_{q'}]_L^{\lambda=L-1}
+
\rho_q\sum_{q'}(b_{qq'}-\frac{L+1}{2L+1}c_{qq'}) [\delta\vec{\nabla}\rho_{q'}]_L^{\lambda=L+1}
\right\}
$
\\
&\\
$[\delta 2i\vec{j}_q]_L^{\lambda=L-1}$ & 
$
-2\times\frac{2m_q^*}{\hbar^2}\left\{
\frac{L}{2L+1}\rho_q \sum_{q'}b_{qq'}[\delta 2i\vec{j}_{q'}]_L^{\lambda=L-1}
-
\frac{\sqrt{L(L+1)}}{2L+1}\rho_q \sum_{q'}b_{qq'}[\delta 2i\vec{j}_{q'}]_L^{\lambda=L+1}
\right\}
$
\\
&\\
$[\delta 2i\vec{j}_q]_L^{\lambda=L+1}$ & 
$
2\times\frac{2m_q^*}{\hbar^2}\left\{
\frac{\sqrt{L(L+1)}}{2L+1}\rho_q\sum_{q'} b_{qq'}[\delta 2i\vec{j}_{q'}]_L^{\lambda=L-1}
-
\frac{L+1}{2L+1}\rho_q \sum_{q'} b_{qq'}[\delta 2i\vec{j}_{q'}]_L^{\lambda=L-1}
\right\}
$
\\
&\\
$[\delta\tau_q]_L$ &
$
2\times\frac{2m^*_q}{\hbar^2}
\left\{
\tau_q\sum_{q'}b_{qq'}[\delta\rho_{q'}]_{L}
+
\frac{1}{2}
\del{\rho_q}{r}\sum_{q'}(b_{qq'}-c_{qq'})
\left(
\sqrt{\frac{L}{2L+1}}[\delta\vec{\nabla}\rho_{q'}]_L^{\lambda=L-1}
-\sqrt{\frac{L+1}{2L+1}}[\delta\vec{\nabla}\rho_{q'}]_L^{\lambda=L+1}\right)
\right\}
$
\\
&\\
\hline
\end{tabularx}
\caption{The second term of r.h.s. of the linear response equation(\ref{eq:lineq2})}
\label{tb:appendix}
\end{table}
\end{center}
\end{widetext}

\bibliographystyle{revtex}
\bibliography{manuscript_revised}

\end{document}